\documentclass[twocolumn,showpacs,preprintnumbers,amsmath,amssymb,APSl,prd,nofootinbib,superscriptaddress]{revtex4-2}
\usepackage{graphicx,color}
\usepackage{subfig}
\usepackage{amssymb}
\usepackage{hyperref}
\usepackage[utf8]{inputenc}
\usepackage[english]{babel}
\usepackage{epsfig}
\usepackage{wasysym}
\usepackage{color,xcolor}
\usepackage{amsmath}
\usepackage{bm}
\usepackage{epsfig}
\usepackage{amsfonts}
\usepackage{dcolumn}
\usepackage{float}
\usepackage{tikz}
\hypersetup{
	colorlinks=true,
	urlcolor= magenta,
	citecolor=blue,
	linkcolor= blue,
	bookmarks=true,
	bookmarksopen=false,
}
\usetikzlibrary{arrows} 
\usetikzlibrary{decorations.markings}
\usepackage{orcidlink}

\begin{document}

\title{ Topologically Charged Holonomy corrected Schwarzschild black hole lensing}

\author{A. R. Soares\orcidlink{0000-0003-1871-2068}}
\email{adriano.soares@ifma.edu.br }
\affiliation{Instituto Federal de Educaç\~ao Ci\^encia e Tecnologia do Maranh\~ao, Campus Buriticupu, CEP 65393-000, Buriticupu, Maranh\~ao, Brazil.}

\author{R. L. L. Vit\'oria \orcidlink{0000-0001-8802-3634}}
\email{ricardo-luis91@hotmail.com}
\affiliation{Faculdade de F\'isica, Universidade Federal do Par\'a, Av. Augusto Corr\^ea, Guam\'a, 66075-110, Bel\'em, PA, Brazil.}

\author{C. F. S. Pereira \orcidlink{0000-0001-6913-0223}}
\email{carlos.f.pereira@edu.ufes.br}
\affiliation{Departamento de F\'isica e Qu\'imica, Universidade Federal do Esp\'irito Santo, Av.Fernando Ferrari, 514, Goiabeiras, Vit\'oria, ES 29060-900, Brazil}
    

\begin{abstract}
In this paper, we theoretically investigate the deflection of light produced by a topologically charged Holonomy corrected Schwarzschild black hole. The study is carried out both in the weak field limit and in the strong field limit. We analytically deduced the expansions for light deflection in the two limits and, from them, we determined the observables in order to provide elements so that observational tools are able to identify these solutions. We model possible gravitational scenarios in order to verify the possible gravitational characteristics of the solution.
\end{abstract}


\maketitle

\section{Introduction}
\begin{figure}[h]
	\centering
	\includegraphics[scale=0.5]{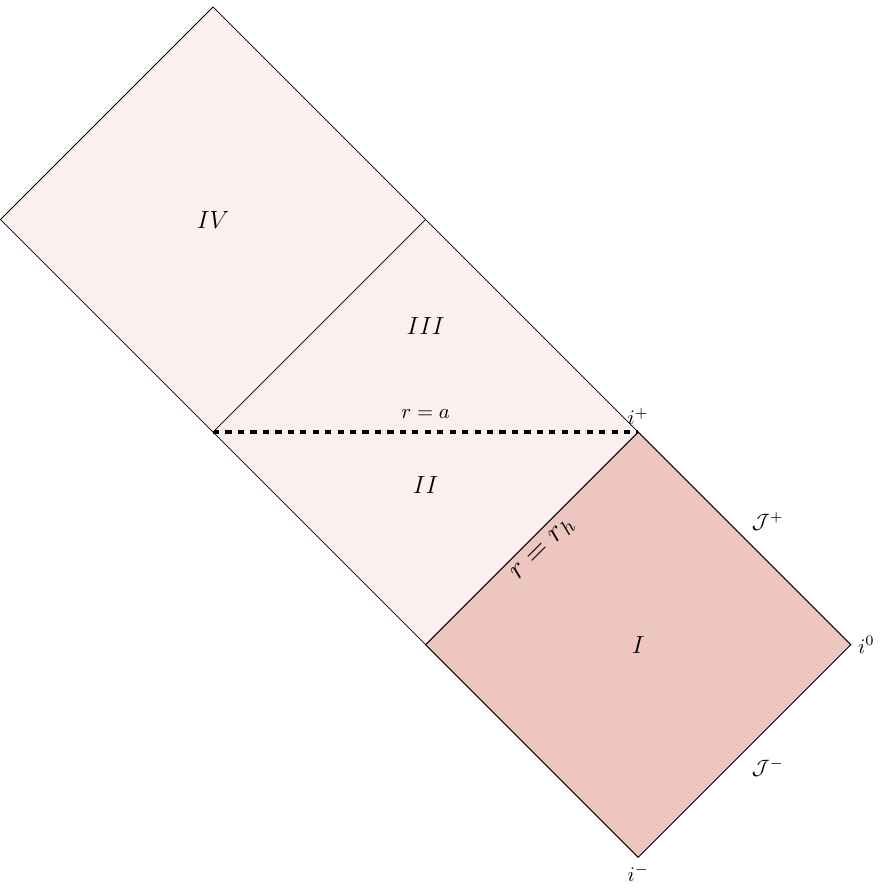}
	\caption{The region I is asymptotically flat and lies beyond the hypersurface $r=r_{h}$. Region II corresponds to the black hole, where $r=a$ defines a minimal spacelike hypersurface separating the trapped regular BH interior from the anti-trapped other white-hole region III. And IV is another asymptotically flat region.}
	\label{pen}
\end{figure}

Despite the great scientific and technological advances provided by General Relativity (GR) \cite{tst}, this theory has problems with geodesic singularities, as is the case with black holes and the big bang \cite{1457,529}. Faced with this scenario, among other cosmological issues \cite{SSTC, SCPC, Knop, BoomerangCollaboration, Steinhardt, Halverson}, physicists have been working on alternative gravitational theories that are consistent with current observations and that are capable of avoiding geodesic singularities. Among these theories, we highlight Loop Quantum Gravity (LQG) \cite{Rovelli003, Ashtekar84, Perez80}. LQG is a non-perturbative theory for quantizing the structure of spacetime and, although it does not yet present a complete quantum description close to a singularity, it has presented effective models in low-energy regimes with corrections arising from quantum effects. Recently,
in \cite{Alonso-Bardaji829, Alonso-Bardaji106},
using LQG, the authors derived a spacetime solution corresponding to a singularity-free interior (black hole/white hole) and two asymptotically flat outer regions. The inner region contains a black-bounce surface, replacing the standard Schwarzschild spacetime singularity. The authors found the global causal structure and the maximum analytical extension, as illustrated in the diagram in Fig.\ref{pen}.
The metric line element that describes the holonomy corrected Schwarzschild black hole in region I of the diagram in Fig.\ref{pen}, in spherical coordinates ($t,r,\theta, \phi$), is given by \cite {Zhang08421, Moreira107}
\begin{eqnarray}\label{me}
	ds^2&=&-\bigg(1-\frac{2M}{r}\bigg)dt^2+\frac{r}{r-a}\bigg(1-\frac{2M}{r}\bigg)^{-1}dr^2\nonumber\\
	&& + r^2\left(d\theta^2+\sin^2\theta d\phi^2\right) \ ,
\end{eqnarray}
where $a$ is the LQG parameter with $a<2M$. Many aspects of this spacetime have already been investigated in order to provide possible observational signatures, enabling parameters that indicate the plausibility of LQG theory, we can mention those linked to quasinormal modes \cite{Zhang08421, Moreira107}, horizon area \cite{ Sobrinho14} and gravitational lensing \cite{LQg-lesing, Ednaldo}. Let us, in a complementary way, draw attention to the fact that such a solution, (\ref{me}), has implications in other  research areas, in particular, those linked to possible phase transitions in the primordial universe. It is theoretically known that these transitions may have given rise to topological defects such as the Global Monopole (GM), resulting from the following pattern of spontaneous symmetry breaking: $SO(3)\times U(1)$ \cite{Kibble, vil}. The gravitational field generated by this defect was originally studied by Barriola and Vilenkin, \cite{Barriola-Vilenkin}, and presents a topological charge \footnote{The   topological charge, $Q$, origin is the GM model and can be calculated following its definition in Ref.\cite{vil}, $ Q=\frac{1}{8\pi}\oint dS^{ij}|\phi|^{-3}\varepsilon_{a b c}\phi^a \partial_{i}\phi^b\partial_{j}\phi^c$, and the matter source in \cite{Barriola-Vilenkin}.}. One of its main characteristics is the angular deficit that influences the geodesics linked to this spacetime. We must highlight that, taking into account the LQG, the gravitational field of the GM presents new characteristics that imply observational signatures different from those already known for the standard GM. In this sense, in \cite{armed2024}, a spacetime type (\ref{me}) was theorized, but now with  GM, whose metric, in spherical coordinates, is given by
\begin{eqnarray}\label{me2}
	ds^2&=&-\bigg(1-\alpha^2-\frac{2M}{r}\bigg)dt^2\nonumber\\
	&&+\frac{r}{r-a}\bigg(1-\alpha^2-\frac{2M}{r}\bigg)^{-1}dr^2\nonumber\\
	&& + r^2\left(d\theta^2+\sin^2\theta d\phi^2\right) \ ,
\end{eqnarray}
where $\alpha^2$ is a dimensionless parameter linked to the energy of spontaneous symmetry breaking \footnote{For a typical unification scale, this term is actually very small: $\sim 10^{-5}$, \cite{Barriola-Vilenkin}.} and $a<\frac{2M}{1-\alpha^2}$. In \cite{armed2024}, among other issues, the author investigated the null geodesics of the system in the weak field limit and showed, in the case of the standard GM, how the deflection of light is amplified by the presence of the GM parameter. However, the author neglected lensing in the strong field limit, which is precisely where most alternative theories of gravity predict results significantly different from those predicted by Einstein's Relativity.
Considering that lensing, in both limits, has already been studied in the Schwarzschild spacetime with topological charge \cite{Pramana, Man2011} and in the Holonomy corrected Schwarzschild spacetime \cite{LQg-lesing}, we conclude that it is  important to carry out a complete study of lensing in the Holonomy corrected Schwarzschild black hole with topological charge (GM). In this sense, we propose a complete and comparative study with the scenarios already presented in the literature.

The work is divided as follows: In Sec.\ref{secII}, we obtain the null geodesic equations and the deflection of light in the weak field limit. In Sec.\ref{secIII}, we analytically calculate the deflection of light in the strong field limit. In Sec.\ref{secIV} we briefly review lensing and study the observables. Finally, we conclude in Sec.\ref{secV}, reviewing the results of the work.

\section{Geodesic Equations and Lensing}\label{secII}
Gravitational lensing consists of the deflection of light when propagating in a gravitational field. This phenomenon, which showed for the first time that GR adequately describes gravitational phenomena \cite{varios3}, has become an important research tool in cosmology and astrophysics, contributing to topics such as the distribution of structures \cite{Mellier1999, Schneider2001}, dark matter \cite{Kaiser1993}, black holes \cite{Eiroa2002, Eiroa2004, Tsukamoto2017, rotacao, Aazami, azquez, Bozza-Sereno, Bozza-Scarpetta, VBz, soaresBbounce, ARC2023Lesing}, wormholes \cite{me2, Nakajima85, Chetouani, Nandi, Dey-sen, Bhattacharya, Gibbons-Vyska, Tsukamoto-Harada, Nandi-Potapov, Tsukamoto2017-95, Tsukamoto-Harada-2017, Shaikh-Banerjee}, topological defects \cite{Cheng, Sharif2015, Cheng92, Cheng28}, theories modified gravity \cite{Sotani92, Wei75, Bhadra, Eiroa73, Sarkar23, Mukherjee39, Gyulchev75, Chen80, Shaikh96, EibIdefle, Paulo2024arxiv} and regular black holes \cite{Eiroa28, Eiroa88}. Lensing can be divided into two regimes. The first is called the weak field limit, when the light passes very far from the gravitational object that originates the lens and the strong field limit, when the light passes very close to the gravitational object so that the angular deflection is divergent at a certain limit approach. In order to study the two limits, let us next obtain the geodesic equations.

\subsection{Geodesic Equations}
 For a smooth curve on a space with metric (\ref{me2}), the lenght, $S$, of tha curve is
\begin{equation}\label{eq33}
	S=\int \sqrt{\bigg(g_{\mu\nu}\frac{dx^\mu}{d\lambda}\frac{dx^\nu}{d\lambda}\bigg)}d\lambda \ ,
\end{equation}
where $\lambda$ is the affine parameter of the curve. Taking $S$ as the affine parameter itself in (\ref{eq33}), we can show that the curves that minimize (\ref{eq33}), $\delta S=0$, also minimize:

\begin{equation}
	\int \bigg(g_{\mu\nu}\frac{dx^\mu}{d\lambda}\frac{dx^\nu}{d\lambda}\bigg)d\lambda \ =\int \mathcal{L} d\lambda \ .
\end{equation}
Therefore, for $\theta = \frac{\pi}{2}$, the Lagrangian $\mathcal{L}$ becomes:
\begin{eqnarray}\label{l}
	\mathcal{L}&=&-\bigg(1-\alpha^2-\frac{2M}{r}\bigg)\bigg(	\frac{dt}{d\lambda}\bigg)^2\nonumber\\ &&+\frac{r}{r-a}\bigg(1-\alpha^2-\frac{2M}{r}\bigg)^{-1}\bigg(\frac{dr}{d\lambda}\bigg)^2\nonumber\\
	 &&+r^2\bigg(\frac{d\phi}{d\lambda}\bigg)^2 \ .
\end{eqnarray}
The corresponding Euler-Lagrange equation for the coordinates $t$ and $\phi$ leads to the following conserved quantities
\begin{equation}\label{en}
	E=\bigg(1-\alpha^2-\frac{2M}{r}\bigg)\bigg(\frac{dt}{d\lambda}\bigg) \ ,
\end{equation}
and
\begin{equation}\label{mom}
	L=r^2\frac{d\phi}{d\lambda} \ .
\end{equation}
which can be understood as energy and angular momentum. Replacing (\ref{en}) and (\ref{mom}) into (\ref{l}) and considering null geodesics, where $\mathcal{L}=0 $, the (\ref{l}) leads to

\begin{equation}\label{eq1}
	\frac{r}{r-a}\bigg(\frac{dr}{d\lambda}\bigg)^2=E^2-\frac{L^2}{r^2}\bigg(1-\alpha^2-\frac {2M}{r}\bigg) \ .
\end{equation}
Eq.(\ref{eq1}) can be seen as describing the dynamics of a classical particle of energy $E$ subject to an effective potential
\begin{equation}\label{Vef}
	V_{eff}=\frac{L^2}{r^2}\bigg(1-\alpha^2-\frac{2M}{r}\bigg) \ .
\end{equation}
To find the radius of the photon sphere, $r_m$, it is enough to take $\frac{dV_{eff}(r)}{dr}=0$, doing this for the topologically
holonomy corrected Schwarzschild BH, we found
\begin{equation}\label{eq2}
	r_m= \frac{3M}{1-\alpha^2} \ ,
\end{equation}
as occurs in Schwarschild spacetime with topological charge \cite{Man2011}. The radius of the photon sphere, Eq.\ref{eq2}, constitutes a fundamental parameter in our study, as we will consider the movement of light far away (weak field limit) and close to it (strong field limit).

\subsection{Expansion for Light deflection in the weak field limit}
In order to introduce the elements for calculating the deflection of light, let us consider a photon starting from the asymptotically flat region and approaching the BH at a radial distance $r_{0}$ from the center of the BH, called the turning point , such that $r_{0}>r_m$. After being deflected by the BH's gravitational field, the photon heads to another asymptotically flat region. At the turning point, we have $V_{eff}(r_{0})=E^2$, which leads to the following expression
\begin{equation}\label{eq3}
	\frac{1}{\beta^2}=\frac{1}{r_{0}^2}\bigg(1-\alpha^2-\frac{2M}{r_{0}}\bigg) \ .
\end{equation}
Where $\beta(r_{0})=\frac{L}{E}$ is the impact parameter. Replacing (\ref{mom}) in (\ref{eq1}), we find
\begin{equation}\label{eq4}
	\Big(\frac{d\phi}{dr}\Big)=\bigg[\bigg(1-\frac{a}{r}\bigg)\bigg[\frac{r^4}{\beta^2}-r^2\bigg(1-\alpha^2-\frac{2M}{r}\bigg)\bigg]\bigg]^{-1/2} \ .
\end{equation}
 We want to find the change in coordinate $\phi$, i.e., $\Delta\phi= \phi_{-}-\phi_{+}$.  By symmetry, the contributions to $\Delta\phi$ before and after the turning point are equal, so
Eq.(\ref{eq4}) leads to
\begin{eqnarray}\label{eq55}
	\Delta\phi&=&2\int_{r_{0}}^{\infty}\bigg[\bigg(1-\frac{a}{r}\bigg)\bigg[\frac{r^4}{\beta^2}\bigg.\nonumber\\
	\bigg.&&-r^2\bigg(1-\alpha^2-\frac{2M}{r}\bigg)\bigg]\bigg]^{-1/2} dr \ .
\end{eqnarray}
Introducing the following variable change $u=\frac{1}{r}$, from which, we have $dr=-\frac{du}{u^2}$. Furthermore, $u\to0$ when $r\to\infty$ and $u\to u_0$ when $r\to r_0$. Therefore, in terms of $u$, (\ref{eq55}) becomes
\begin{equation}\label{eq5}
	\Delta\phi=2\int_{0}^{u_{0}}\bigg[(1-au)\bigg[\frac{1}{\beta^2}-u^2(1-\alpha^2-2Mu)\bigg]\bigg]^{-1/2} du\ .
\end{equation}
From (\ref{eq3}), we have $1/\beta^2=u_o^2(1-\alpha^2-2Mu_o)$, which substituting in (\ref{eq5}), implies
 \begin{eqnarray}\label{eq6}
	\Delta\phi&=&2\int_{0}^{u_{0}}\bigg[(1-au)\bigg[ u_0^2(1-\alpha^2-2Mu_0)\bigg.\nonumber\\	
&&\bigg.-u^2(1-\alpha^2-2Mu)\bigg]\bigg]^{-1/2} du\ .
\end{eqnarray}

In the weak field approximation, that is, assuming that the photon passes very far from the BH, we can take the approximation $M\ll1$ and $a\ll1$. Therefore, up to second order in $a$, a (\ref{eq6}) provides the deflection of light $\delta\phi=\Delta\phi-\pi$:
\begin{eqnarray}\label{eq7}
	\delta\phi&\simeq&\left(\frac{1}{\sqrt{1-\alpha^2}}-1\right)\pi+\frac{4M}{\beta(1-\alpha^2)^{3/2}} \nonumber\\
	&+&\frac{a}{\beta\sqrt{1-\alpha^2}}+\frac{3\pi a^2}{16\beta^2\sqrt{1-\alpha^2}}\nonumber\\
	&+&\frac{aM(3\pi-4)}{4\beta^2(1-\alpha^2)^{3/2}} \ .
\end{eqnarray}
In Eq.(\ref{eq7}), the first two terms refer to the deflection in the standard Schwarszchild BH spacetime, without the effects of the holonomic correction; for $\alpha$ small, as indeed it must be, it reduces to
\begin{equation}
	\delta\phi=\frac{4M}{\beta(1-\alpha^2)^{3/2}} \ ,
\end{equation}
which is in agreement with \cite{Pramana}. The following terms, in (\ref{eq7}), bring a contribution from the Holonomic correction. Later, we will study the observational implications of these corrections.

 \section{Deflection of light in the strong field limit}\label{secIII}
 To derive the deflection of light in the strong field limit, we will adopt the methodology developed by Bozza \cite{Bozza2002} and improved by Tsukamoto \cite{Tsukamoto2017}.
 
Making the following variable change
 \begin{equation}\label{eq8}
 	z=1-\frac{r_0}{r} \ ,
 \end{equation}
 the Eq.(\ref{eq55}) becomes
 \begin{equation}\label{eq9}
 	\Delta\phi(r_0)=\int_{0}^{1} \frac{2r_0}{\sqrt{G(z,r_0)}}\ dz \ ,
 \end{equation}
 where,
 \begin{eqnarray}\label{eq10}
 	G(z,r_0)&=&\frac{r_0^4}{\beta^2}-\frac{ar_0^3}{\beta^2}(1-z)-(1-\alpha^2)r_0^2(1-z)^2 \nonumber\\
 	&&+(2M+a(1-\alpha^2))r_0(1-z)^3\nonumber\\
 	&&-2Ma(1-z)^4 \ .
 \end{eqnarray} 
Expanding $G(z,r_0)$ in a power series close to $z=0$ (which corresponds to $r\to r_0$), we get
\begin{equation}\label{eq11}
	G(z,r_0)\simeq\Lambda_1(r_0)z+\Lambda_2(r_0)z^2 \ ,
\end{equation}
where
\begin{equation}\label{eq12}
	\Lambda_1(r_0)=2(r_0-a)\left[(1-\alpha^2)r_0-3M\right]
\end{equation}
and 
\begin{equation}\label{eq13}
	\Lambda_2(r_0)=6M(r_0-2a)-(1-\alpha^2)r_0(r_0-3a) \ .
\end{equation}
In the strong field limit, when $r_0\to r_m=3M/(1-\alpha^2)$, the expansion coefficients become
\begin{equation}\label{eq14}
	\Lambda_1(r_0)\to\Lambda_1(r_m)=0 
\end{equation}
and
\begin{equation}\label{eq141}
	\Lambda_2(r_0)\to\Lambda_2(r_m)=\frac{9M^2}{(1-\alpha^2)}-3aM \ .
\end{equation}
The equations (\ref{eq14}) and (\ref{eq141}) show that in the strong field limit
the integral (\ref{eq9}) diverges logarithmically. In order to obtain an expression for the deflection of light in this limit, we will divide Eq.(\ref{eq9}) into two parts, a divergent part $\Delta\phi_D(r_0)$ and a regular part $\Delta\phi_R(r_0)$, so that
\begin{equation}\label{eq15}
	\Delta\phi_R(r_0)=\Delta\phi(r_0)-\Delta\phi_D(r_0) \ .
\end{equation}
The divergent part is given by
\begin{equation}
	\Delta\phi_D(r_0)=\int_{0}^{1} \frac{2r_0}{\sqrt{\Lambda_1(r_0)z+\Lambda_2(r_0)z^2}}\ dz \ ,
\end{equation}
whose integration provides
\begin{eqnarray}\label{eq16}
	\Delta\phi_D(r_0)&=&-\frac{4r_0}{\sqrt{\Lambda_2(r_0)}}\log(\sqrt{\Lambda_1(r_0)})\nonumber\\
	&& +\frac{4r_0}{\sqrt{\Lambda_2(r_0)}}\log\bigg(\sqrt{\Lambda_2(r_0)}\bigg.\nonumber\\
	&&\bigg.+\sqrt{\Lambda_1(r_0)+\Lambda_2(r_0)}\bigg) \ .
\end{eqnarray}
 Expanding $\Lambda_ 1(r_ 0)$ and $\beta(r_ 0)$ close to the photosphere radius, $r_m$, of (\ref{eq12}) and (\ref{eq3}), we obtain
\begin{equation}\label{eq17}
	\Lambda_1(r_0)\simeq(6M-2a(1-\alpha^2))\left(r_0-\frac{3M}{1-\alpha^2}\right) \ 
\end{equation}
and
\begin{eqnarray}\label{eq18}
	\beta(r_0)&\simeq&\sqrt{\frac{27M^2}{(1-\alpha^2)^3}}\nonumber\\
	&&+\sqrt{\frac{3(1-\alpha^2)}{4M^2}}\left(r_0-\frac{3M}{1-\alpha^2}\right)^2
\end{eqnarray}
From (\ref{eq17}) and (\ref{eq18}), we have
\begin{eqnarray}\label{eq19}
	\Lambda_1(r_0)&\simeq&2\sqrt{6}\left(\frac{3M^2-a(1-\alpha^2)M}{1-\alpha^2}\right)\nonumber\\
	&&\times
	\sqrt{\bigg(\frac{\beta(1-\alpha^2)^{3/2}}{\sqrt{27M^2}}-1\bigg)} \ .
\end{eqnarray}
Replacing (\ref{eq19}) in (\ref{eq16}) and considering the strong field limit, that is, $r_0\to r_m=3M/(1-\alpha^2)$, we get

\begin{eqnarray}\label{eq20}
	\Delta\phi_D&=&-\sqrt{\frac{3M}{(1-\alpha^2)(3M-(1-\alpha^2)a)}}\nonumber\\	
	&&\times\log\bigg(\frac{\beta}{M}\left(\frac{1-\alpha^2}{3}\right)^{3/2}-1\bigg) \nonumber\\
	&&+\sqrt{\frac{3M}{(1-\alpha^2)(3M-(1-\alpha^2)a)}}\log(6) \ .
\end{eqnarray}
It is worth noting that we can easily revisit the cases already discussed in the literature taking appropriate limits. For example, taking $a=0$, that is, without correction from LQG, we fall back on the result obtained in \cite{Man2011}. And if we take $\alpha=0$, that is, without a global monopole, we fall back on the result obtained in \cite{LQg-lesing}.

The regular part, (\ref{eq15}), is given by
\begin{eqnarray}\label{eq21}
	\Delta\phi_R(r_0)&=&\int_{0}^{1} \frac{2r_0}{\sqrt{G(z,r_0)}}\ dz \nonumber\\
	&&-\int_{0}^{1} \frac{2r_0}{\sqrt{\Lambda_1(r_0)z+\Lambda_2(r_0)z^2}}\ dz \ .
\end{eqnarray}

In the strong field limit, $r_0\to r_m=3M/(1-\alpha^2)$, the impact parameter given by Eq.(\ref{eq3}) tends to a critical value $\beta_c=\beta (r_m)=\left(\frac{3}{1-\alpha^2}\right)^{3/2}M$. Considering the equations (\ref{eq10}), (\ref{eq11}) and (\ref{eq3}), (\ref{eq21}) can be written as
\begin{eqnarray}\label{eq22}
	\Delta\phi_R&=&\int_{0}^{1}2\left\{\frac{1-\alpha^2}{3}  -\frac{a(1-\alpha^2)^2}{9M}(1-z)\right.\nonumber\\
	&&\left. -(1-\alpha^2)(1-z)^2-\frac{2a(1-\alpha^2)^2}{9M}(1-z)^4   \right.\nonumber\\
	&&\left.+\frac{(1-\alpha^2)(2M+a(1-\alpha^2))}{3M}(1-z)^3 \right\}^{-1/2}dz\nonumber\\
	&&-\int_{0}^{1}2\left\{(1-\alpha^2)-\frac{(1-\alpha^2)^2a}{3M}\right\}^{-1/2}z^{-1}dz
\end{eqnarray}
Despite its unpleasant appearance, the integral (\ref{eq22}) produces an exact value for $a\le\frac{3M}{1-\alpha^2}$, which is within the limit we are assuming. Integrating (\ref{eq22}), we obtain
\begin{widetext}
	\begin{equation} \label{eq23}
			\Delta\phi_R=\sqrt{\frac{12}{(1-\alpha^2)(3-2(a(1-\alpha^2)/2M))}} 
			\log\bigg(\frac{18-12(a(1-\alpha^2)/2M)}{6-(a(1-\alpha^2)/2M)+3\sqrt{3-2(a(1-\alpha^2)/2M)}}\bigg) \ .
		\end{equation}
\end{widetext}

Therefore, from (\ref{eq15}), joining (\ref{eq20}) and (\ref{eq23}), we finally explicitly find the expansion for the deflection of light in the strong field limit, $\delta\phi =\Delta\phi-\pi$,
\begin{widetext}
	\begin{eqnarray}\label{eq24}
		\delta\phi&=&-\sqrt{\frac{3M}{(1-\alpha^2)(3M-(1-\alpha^2)a)}}	
		\log\bigg(\frac{\beta}{M}\left(\frac{1-\alpha^2}{3}\right)^{3/2}-1\bigg) +\sqrt{\frac{3M}{(1-\alpha^2)(3M-(1-\alpha^2)a)}}\log(6)\nonumber\\
			&&+\sqrt{\frac{12}{(1-\alpha^2)(3-2(a(1-\alpha^2)/2M))}} 
			\log\bigg(\frac{18-12(a(1-\alpha^2)/2M)}{6-(a(1-\alpha^2)/2M)+3\sqrt{3-2(a(1-\alpha^2)/2M)}}\bigg)	-\pi \ .	
	\end{eqnarray}
\end{widetext}
Taking the appropriate limits in (\ref{eq24}), we easily recover the deflection in the cases already discussed in the literature. For example, taking $a=0$, we fall back on the result obtained in \cite{Man2011}, that is, with GM and without LQG effects. And if we take $\alpha=0$, that is, without GM, we fall back on the results obtained in \cite{LQg-lesing}.

\begin{figure}[h]
	\centering
	\includegraphics[width=\columnwidth]{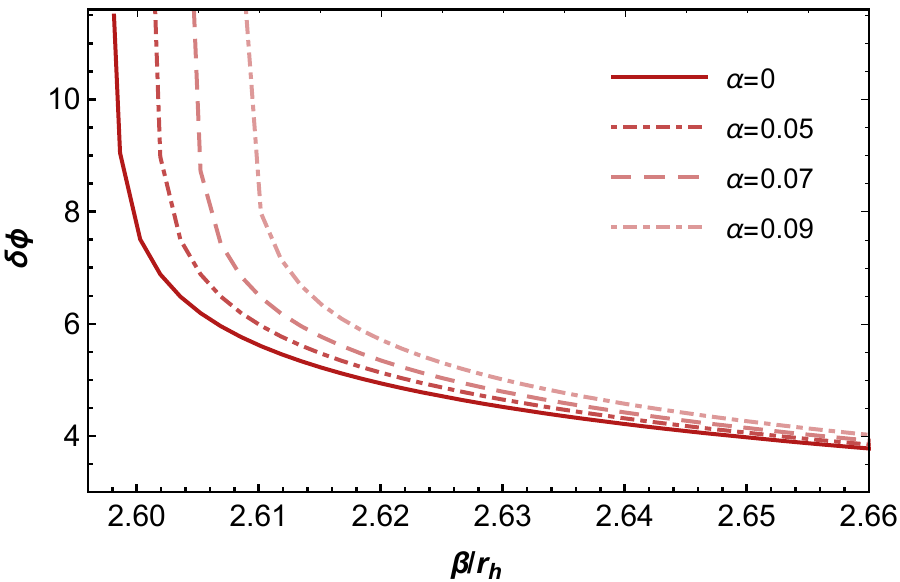}
	\caption{Light angular deflection  as a function de $\beta/r_h$ para $a/r_h=0.3$ .} 
	\label{LG}
\end{figure}

In Fig.\ref{LG}, plotamos a deflexão da luz para alguns valores de $\alpha$ taking a specific value for the ratio between the LQG parameter and the radius of the event horizon, $r_h=\frac{(1-\alpha^2)}{2M}$, of the solution (\ref{me2}), in order to graphically show that the presence of the GM amplifies the deflection.

 \section{Lens Equation}\label{secIV}
In this section, we will substitute the expressions found for the deflection of light, (\ref{eq7}) and (\ref{eq24}), into the lens equations to generate, theoretically, quantities that can be useful observationally, that is, that allow verify the existence of the solution studied in this work and distinguish it from the Schwarzschild black hole. Therefore, let us first briefly review the lens equations in the strong field limit.

\begin{figure}[h]
	\centering
	\includegraphics[width=\columnwidth]{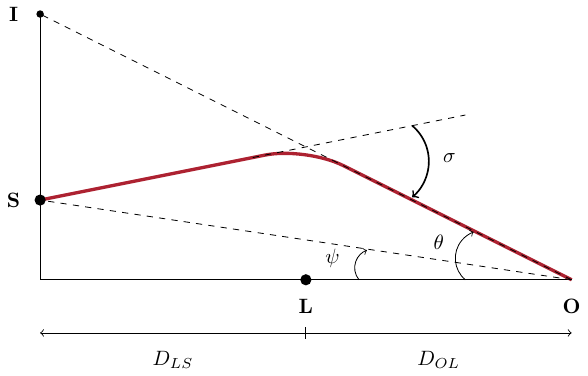}
	\caption{Light angular deflection  diagram} 
	\label{fig3}
\end{figure}

In Fig. (\ref{fig3}), we visually diagram the lensing. The light that is emitted by the source $S$ is deflected towards the observed $O$ by the LQG compact object with topological charge located in $L$. The angular deflection of light is given by $\sigma$. The angular positions of the source and image in relation to the optical axis, $\overline{LO}$, are given, respectively, by $\psi$ and $\theta$. 
 
Let us admit that the source ($S$) is almost perfectly aligned with the lens ($L$) which is where relativistic images are most expressive, \cite{Boz-Cap2001,Virbhadra-Ellis-2000}. Therefore, the lens equation relating the angular positions $\theta$ and $\psi$ is given by
\begin{equation}\label{EqLente}
	\psi=\theta-\frac{D_{LS}}{D_{OS}}\Delta\sigma_{n}\ ,
\end{equation}
where $\Delta\sigma_n$ is the deflection angle subtracted from all the loops made by the photons before reaching the observer, that is, $\Delta\alpha_{n}=\alpha-2n\pi$. In this approximation, from very small angular positions,
\begin{equation}
	\beta\simeq\theta D_{OL} \ .
\end{equation}
See that the angular deflection (\ref{eq24}) can be written as
\begin{equation}\label{defle}
	\sigma(\theta)=-\bar{a}\log\left(\frac{\theta D_{OL}}{\beta_c}-1\right)+\bar{b}\ ,
\end{equation}
where,
\begin{equation}
	\bar{a}= \sqrt{\frac{3M}{(1-\alpha^2)(3M-(1-\alpha^2)a)}}	\ ,
\end{equation}
\begin{eqnarray}
	\bar{b}&=&\sqrt{\frac{3M}{(1-\alpha^2)(3M-(1-\alpha^2)a)}}	\log(6)\nonumber\\
	&& +\Delta\phi_R-\pi \ .
\end{eqnarray}
and
\begin{equation}
	\beta_c=\left(\frac{3}{1-\alpha^2}\right)^{3/2}M \ .
\end{equation}
See that the expression for $\Delta\phi_R$ is given in Eq.(\ref{eq23}).
What enters the lens equation is $\Delta\sigma_{n}$, to obtain it we expand $\sigma(\theta)$ close to $\theta=\theta^{0}_n$, where $\sigma (\theta^{0}_n)=2n\pi$. Thus, we are left with
\begin{equation}\label{da}
	\Delta\sigma_n=\frac{\partial\sigma}{\partial\theta}\Bigg|_{\theta=\theta^0_n}(\theta-\theta^0_n) \ .
\end{equation}
Evaluating (\ref{defle}) in $\theta=\theta^{0}_n$, we obtain
\begin{equation}\label{To}
	\theta^0_{n}=\frac{\beta_c}{D_{OL}}\left(1+e_n\right), \qquad\text{where}\quad e_n=e^{\frac{\bar{b}-2n\pi}{\bar{a}}} \ .
\end{equation}
Substituting (\ref{defle}) and (\ref{To}) into (\ref{da}), we get
\begin{equation}\label{de}
	\Delta\sigma_n=-\frac{\bar{a}D_{OL}}{\beta_ce_n}(\theta-\theta^0_n) \ .
\end{equation}
Substituting (\ref{de}) in the lens equation (\ref{EqLente}), we obtain the expression for the $n$th angular position of the image
\begin{equation}\label{imag}
	\theta_n\simeq\theta^0_n+\frac{\beta_ce_n}{\bar{a}}\frac{D_{OS}}{D_{OL}D_{LS}}(\psi-\theta^0_n) \ .
\end{equation}

 The total flux received by a lensed image is proportional to the magnification $\mu_{n}$, which is given by $	\mu_n=\left|\frac{\psi}{\theta}\frac{\partial\psi}{\partial\theta}|_{\theta^0_{n}}\right|^{-1}$. Then, from (\ref{EqLente}) and (\ref{de}) , we get
\begin{equation}\label{flu}
	\mu_{n}=\frac{e_n(1+e_n)}{\bar{a}\psi}\frac{D_{OS}}{D_{LS}}\left(\frac{\beta_c}{D_{OL}}\right)^2 \ .
\end{equation}
We observe that  $\mu_n$ decreases very  fastly with $n$, so the brightness of the first image $\theta_1$ dominates  in comparing with other ones. In any case, due to the factor $\left(\frac{\beta_c}{D_{OL}}\right)^2$ it is clear that the magnification will always be small.

The expressions for the relativistic images (\ref{imag}) and their respective fluxes (\ref{flu}) were constructed in terms of the expansion coefficients ($\bar{a}$, $\bar{b}$ , and $b_c$). Let us now consider the inverse problem, that is, from observations, determine the expansion coefficients. With this, we can understand the nature of the object that generates the gravitational lens and compare it with the predictions made by the present study.
From (\ref{imag}) and (\ref{To}), taking $n\to\infty$ we conclude that the critical impact parameter can be expressed as
\begin{equation}
	\beta_c=D_{OL}\theta_{\infty} \ ,
\end{equation}
where $\theta_{\infty}$ corresponds to the asymptotic position of the relativistic images.
As in \cite{Bozza2002}, we will assume that only the outermost image $\theta_{1}$ is discriminated as a single image while the others are encapsulated in $\theta_{\infty}$. Therefore, the following observables are defined
\begin{eqnarray}
	s&=&\theta_{1}-\theta_{\infty}= \theta_{\infty} e^{\frac{\bar{b}-2\pi}{\bar{a}}},\\
	\tilde{r}&=& \frac{\mu_{1}}{\sum_{n=2}^{\infty} \mu_{n} }= e^{\frac{2\pi}{\bar{a}}} \ .
\end{eqnarray}
In the expressions above, $s$ is the angular separation and $\tilde{r}$ is the relationship between the flow of the first image and the flow of all others. These forms can be inverted to obtain the expansion coefficients.
 
\subsection{Observables modeled by Sagittarius A*}
In order to verify the impact of the GM on the observables, $\theta_{\infty}, s$ and $\tilde{r}$ we will consider a lensing scenario where the lens is modeled by the black hole at the center of our galaxy, the milky way, \cite{Gezel2010}. Mass is estimated to be $4.4\times10^{6}M_{\odot}$ and distance is approximated to be $D_{OL}=8.5$Kpc. In our calculations, we take into account that, in geometric units, $M\to M\frac{G}{c^2}$. Then, for the modeled scenario, we present the behavior of the observables.

\begin{figure}[h]
 	\centering
 	\includegraphics[width=\columnwidth]{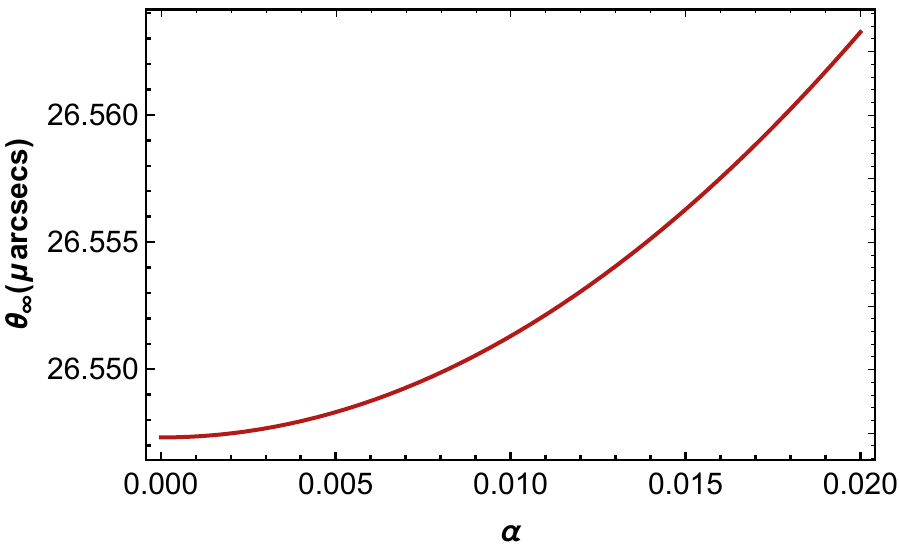}
 	\caption{$\theta_{\infty}$ em função do GM parameter $\alpha$.} 
 	\label{thetainf}
 \end{figure}

We start with $\theta_{\infty}$, whose behavior as a function of the GM parameter, $\alpha$, is plotted in Fig.\ref{thetainf}. In the case without GM, that is, with $\alpha=0$, the asymptotic position of the images is given by $\theta_{\infty}\sim 26.54\mu\text{arcsecs}$. In this sense, the presence of the GM increases the value of $\theta_{\infty}$ compared to other cases.

\begin{figure}[h]
	\centering
	\includegraphics[width=\columnwidth]{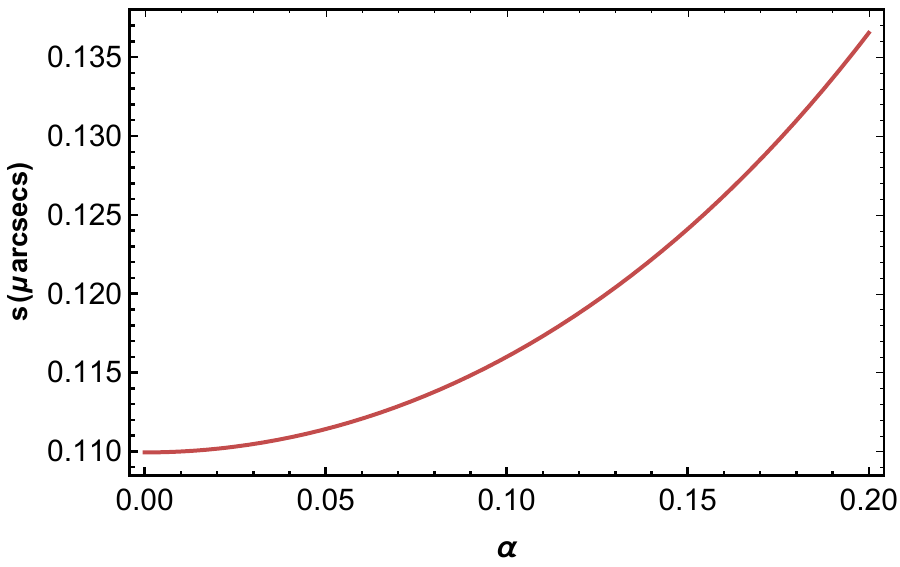}
	\caption{Angular separation $s$. } 
	\label{figs}
\end{figure}

\begin{figure}[h]
 	\centering
 	\includegraphics[width=\columnwidth]{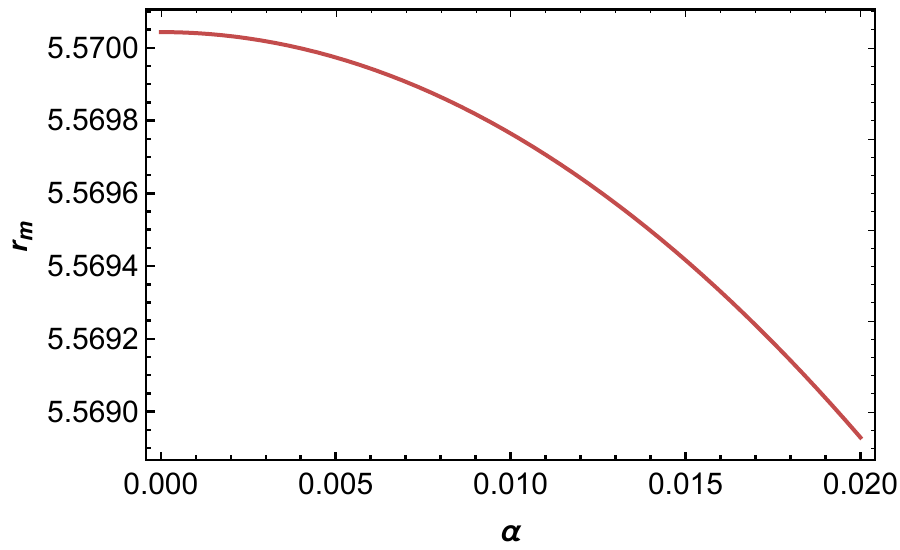}
 	\caption{$	r_m=2.5\log_{10}\tilde{r}$ \ .} 
 	\label{figr}
 \end{figure}

In Fig.\ref{figs} and Fig.\ref{figr}, we plot the angular separation $s$, and $\tilde{r}$ (where we redefine it in terms of a logarithmic scale), respectively, as a function of $\alpha$ for a specific value of the ratio between the LQG parameter $a$ and radius of the event horizon $r_h$, i.e., $\frac{a}{r_{h}}=0.5$ \ .

In Table \ref{tab:table1} we still plot the values of $s$ and $r_m$ for some possible values of $\alpha$. As we can see, the angular separation has an increasing behavior with $\alpha$ while $r_m$ decreases. These results may indicate an increase in the brightness of the relativistic images compared to the first image. The presence of a GM, therefore, implies new characteristics compared to the other cases, LQG and Schwarzschild.
\begin{table}[b]
	\caption{\label{tab:table1}%
		Observables to $a/r_{h}=0.5$
	}
	\begin{ruledtabular}
		\begin{tabular}{lll}
			$\alpha$&
			$s(\mu \textrm{arcsecs})$&
			\multicolumn{1}{c}{$r_m$\textrm{(magnitudes)}}\\
			\colrule
			0 & 0.109 & 5.570 \\
			0.05 & 0.111 & 5.563 \\
			0.1 & 0.115 & 5.542\\
			0.15 & 0.124 & 5.507\\
		\end{tabular}
	\end{ruledtabular}
	
\end{table}
\section{Conclusions}\label{secV}
In this work we investigate the influence of a Global Monopole on gravitational lensing in spacetime motivated by LQG theory. First, we calculate the expressions for the deflection of light in the weak field limits (\ref{eq7}), when the light passes far from the photon sphere, and in the strong field limit (\ref{eq24}), which corresponds to the limit at which light approaches the photon sphere. In both expressions, we can clearly observe that the presence of the GM amplifies the lensing. To show the observational impacts generated by the GM, we studied the observables in a scenario in which the lens is modeled by the black hole at the center of our galaxy (Sagittarius A*). We show that the presence of the GM increases the asymptotic position of the relativistic images and the angular separation, in addition to increasing the brightness of the other images. With the increase in the optical resolution of observational projects and the increasing efforts of international collaborations \cite{Gezel2010}, we hope, in the relatively near future, to accurately discriminate between the different observational models.

\section*{Acknowledgements}

The authors C. F. S. P. and R. L. L. V. would like to thank
	CAPES (Coordenação de Aperfeiçoamento de Pessoal
	de Nível Superior) for financial support.



\end{document}